\documentclass[12pt]{iopart}

\usepackage{graphics}
\usepackage[pdftex]{graphicx}
\usepackage{epstopdf}
\begin{document}
\title{Microscopic Picture of Superfluid $^4$He}

\author{Yongle Yu\dag \ \footnote[7]{Email-address: yongle.yu@wipm.ac.cn} and Hailin Luo \ddag}
\address{\dag\ State Key Laboratory of Magnetic Resonance and Atomic and \\
Molecular Physics,
  Wuhan Institute of Physics and Mathematics, \\Chinese Academy of Science,
 West No. 30 Xiao Hong Shan, Wuchang, \\ Wuhan, 430071, China}
\address{\ddag\ Fujian Institute of Research on the Structure of 
 Matter,\\ Chinese Academy of Sciences, 155 Yangqiao Road West, Fuzhou, 350002, China}

\begin{abstract}

We elucidate the microscopic quantum mechanism
of superfluid $^4$He by uncovering a novel characteristic 
of its many-body energy levels. At temperature below 
the transition point, the system's
low-lying levels exhibit a fundamental grouping behavior, 
wherein each level belongs exclusively to a single group. In a 
superflow state, the system establishes thermal 
equilibrium with its surroundings on a group-specific basis.
Specifically,  the levels of a selected 
group, initially occupied, become thermally populated, while the
remaining groups of levels stay vacant 
due to absence of transitions between groups.
The macroscopic properties of the system, such as 
its superflow velocity and thermal 
energy density, are statistically determined 
by the thermal distribution of the 
occupied group. Additionally, we infer that the thermal energy 
of a superflow has an unusual relationship with flow 
velocity, such that the larger the flow velocity, the smaller the thermal energy.
This relationship is responsible for a range of intriguing phenomena, 
including the mechano-caloric effect and the fountain effect, which highlight a
fundamental coupling between the thermal motion and
hydrodynamic motion of the system.Furthermore, we 
present experimental evidence of a counterintuitive 
self-heating effect in $^4$He superflows, confirming that 
a $^4$He superflow carries significant thermal 
energy related to its velocity.

\end{abstract}
\vspace{2pc}

\vspace{2pc}

\section*{I. Introduction}

The discovery of quantum mechanics represents a paradigm 
shift in the evolution of modern civilization. It not only deepens 
our understanding of physical systems at a fundamental level but also 
drives the development of numerous vital technologies 
that have permeated all aspects of human life.

Quantum mechanics offers a unified framework for describing diverse 
physical systems, ranging from atomic and molecular systems to 
condensed matter systems and beyond, despite its counterintuitive 
features such as quantum entanglement and tunneling. According to
the theory's formulation, the physical states of any given system 
can be characterized by a set of quantum wavefunctions (states) of
the system, and the physical processes associated with the system correspond
 to the transitions between these quantum states.

The field of atomic and molecular physics has achieved remarkable success 
in attaining a high degree of accuracy in some quantitative agreements between 
quantum theory and experimental observations. This can be attributed, in
part, to the relatively small number of particles involved in these systems, 
enabling precise calculations of their quantum properties. However, in condensed
 matter physics, the quantitative agreements between quantum theory and experimental 
data are often less satisfactory, with even qualitative understanding proving 
elusive in some instances. This is primarily due to the quantum many-body problem, 
which is associated with a large number of particles in the corresponding system. 
To tackle this challenge, quantum approximations, such as the mean field or 
single-particle approach, are commonly employed. This approach approximates 
the quantum state of a system using a product of single-particle quantum 
wavefunctions and can explain significant phenomena, such as the distinctions 
between conductors, semiconductors, and insulators.

Despite its widespread use, the single-particle approach has proved 
inadequate in addressing several intriguing quantum phenomena. Notably, 
there are exceptions to this method, such as the Laughlin wavefunction, 
which is used to describe the fractional quantum Hall effect.

Superfluid $^4$He is one of the most fundamental condensed matter 
systems that exhibit a wide variety of unusual quantum behaviors. 
The phenomenon of superfluidity \cite{kapitza, allen}, characterized 
by the absence of dissipation, seems to be someway 
at odds with the second law of thermodynamics, which typically 
states that a closed system will reach a thermal state 
of maximum entropy by converting all the energy of an 
ordered motion into thermal energy. Several phenomena of 
superfluid $^4$He, such as the fountain effect \cite{allenfountain} 
and the mechano-caloric effect \cite{mendelssohn}, reveal an 
intriguing coupling between the hydrodynamic and thermal motions 
of the system. A naturally posed question is whether one can 
understand these unusual behaviors in the direct terms of 
quantum states and quantum transitions among them. Is it 
possible to form a quantum picture of superfluid $^4$He 
that is as clear and straightforward as that of an atomic 
system? The present paper seeks to provide an 
affirmative answer to this question.

We present some general physical arguments that lead 
to a novel grouping behavior of low-lying many-body 
levels of superfluid $^4$He at low temperature, with 
each level belonging exclusively to one single group. 
The system establishes a group-specific thermal 
equilibrium with its surroundings, and depending on the 
group occupancy, it can sustain a macroscopic 
superflow state. Moreover, we demonstrate that the 
grouping of its low-lying levels gives rise to an 
intriguing property, namely, a velocity-dependent 
thermal energy for the superflow. Specifically, as 
the velocity of the superflow increases, the thermal 
energy decreases. This velocity-dependent behavior is 
responsible for the fundamental coupling between the 
hydrodynamic and thermal motions of the system.

There are numerous important theoretical works in 
the literature that provide significant insights 
into various aspects of superfluid $^4$He (see, for 
example, \cite{landau1, landau2, bogoliubov, london1, london2, 
tisza, bloch, leggett, feynman} and the references 
in \cite{tilley}). It is worth noting that the present 
theoretical advancement owes much to the inspirations 
from some of these seminal works.
 
The rest of the paper is organized as follows.
In Section II, we present a critical analysis of the 
two-fluid model for superfluid $^4$He, highlighting a 
fundamental inadequacy in this phenomenological model. 
We also discuss several experimental observations 
that contradict the two-fluid model, emphasizing 
the need for a direct microscopic picture of the system.
In Section III, we illustrate some general properties 
of the quantum states of superfluid $^4$He, with 
particular attention to the grouping behavior 
of its low-lying levels.
Section IV argues that the thermal energy of 
superfluid $^4$He naturally depends on the flow 
velocity due to the properties of the occupied 
group(s) and the properties of Galilean transformation.
In Section V, we report an intriguing self-heating 
phenomenon of $^4$He superflows, providing evidence 
that superflowing $^4$He can carry significant 
thermal energy depending on the flow velocity.
Finally, we provide our conclusions and perspectives 
in Section VI.


\section*{II. Fundamental inadequacy of the two-fluid model and 
its inconsistency with experimental observations}

The two-fluid model of superfluid $^4$He is essential to 
the textbook understanding of numerous behaviors of the system. However, 
this phenomenological model has a fundamental inadequacy that 
poses a significant challenge to its reliability.

The two-fluid model postulates the existence of a superfluid 
component that possesses an exotic characteristic of zero entropy. 
With this postulation, the superfluid component is considered 
a thermodynamic sub-system, allowing for the investigation 
of its temperature. However, maintaining consistency with 
the zero entropy assumption requires this temperature to 
be absolute zero, which raises the fundamental question of 
how this zero-temperature component can coexist with 
its thermal surroundings. This would be more "miraculous" 
than spotting ice in molten steel, and the zeroth law of 
thermodynamics unequivocally rules out such coexistence. 
In contrast, we shall demonstrate subsequently that the 
microscopic mechanism of superfluidity is free from 
unrealistic assumptions of zero entropy and adheres to 
the zeroth law of thermodynamics.

There is another unnatural aspect of the two-fluid model. 
According to this model, the sum of the normal and superfluid 
densities remains constant, which corresponds to the total 
density of all $^4$He atoms in the system. The normal density 
is determined by computing the density of quasiparticles, such 
as phonons and rotons, with each quasiparticle corresponding 
exactly to one $^4$He atom in the total density. By 
subtracting the quasiparticle-mapped atoms, the remaining
 $^4$He atoms constitute the superfluid component. It is 
clear that the concept of two components corresponds to a 
bipartition of $^4$He atoms in the system. However, this 
bipartition is not supported at a fundamental level. In 
accordance with the principles of quantum mechanics, 
the microscopic states of the system are specified by 
a large number of many-body wavefunctions. These 
wavefunctions, denoted by $\psi_\alpha(\mathbf{r}_1, \mathbf{r}_2, ..., \mathbf{r}_N)$, 
can be conveniently taken as the eigenstates of the Hamiltonian 
operator of the system, with $\alpha$ labeling the eigenlevels 
and $N$ denoting the total number of atoms. Within these 
wavefunctions, all the atoms must be treated as an 
indivisible entity, possessing complex intrinsic quantum 
correlations among them due to the inter-atomic 
interactions.

In previous studies, several experimental observations 
have directly contradicted the predictions of the 
two-fluid model. This result is not entirely surprising, 
as the main assumption of the two-fluid model is 
subjective in nature.

Two experimental studies \cite{osborne, androni} have 
investigated the behavior of the meniscus of superfluid $^4$He 
in a rotational bucket. According to the two-fluid model, 
only the normal component should rotate, while the 
superfluid component remains stationary. As a result, 
only the normal component generates a meniscus, and 
its height is expected to be proportional to the 
normal density, leading to an increase with temperature 
and vanishing at low temperatures.

However, the experimental findings are contrary to these 
predictions. Specifically, the height of the meniscus 
remains constant with temperature, and its value 
seems to suggest that the entire system is composed 
of the normal component, independent of temperature. 
Furthermore, the experiment \cite{androni} demonstrates 
that the rotating $^4$He exhibits the thermo-mechanical 
effect, which contradicts the conclusion that the 
system is solely composed of the normal component.

A different set of experiments explores the oscillation and attenuation
 of superflow motion between two containers of superfluid $^4$He within 
the same isothermal enclosure \cite{osc_atkins, osc_picus, osc_manchester, 
osc_seki, osc_glick, osc_hammel, osc_hallock}. By positioning the liquid levels
 in the containers differently, a superflow motion, typically in the form
 of a mobile surface film, can be generated through a gravitational 
potential difference. When the liquid levels are equalized, the inertial in
 the superflow subsequently causes the levels to oscillate about the
 equilibrium position. The attenuation of liquid level oscillations 
is discernible, indicative of the dissipation of gravitational potential energy.

Investigating the temperature dependence of the damping rate is of
 particular interest. According to the two-fluid model, the normal 
component causes the dissipation. Therefore, as the ratio of normal 
component approaches zero in the low temperature limit, the damping
 rate shall also vanish. However, the experimental observations show 
that the damping rate remains finite and relatively 
large in the low temperature limit.

A similar behavior is also observed in superconductors. In an 
experiment study \cite{Q_Turneaure, Q_tinkham}, the $Q$ value of a
 microwave cavity made of superconductor niobium is measured. According 
to the two-fluid model, it is expected that the $Q$ value would increase 
exponentially as the temperature approaches absolute zero. However, 
experimental measurements indicate that the $Q$ value approaches
 a constant value below the temperature of 1.3 $K$.

The dissipation behavior of superfluid $^4$He and superconductors 
is considered an open question in literature. Later in this 
work, we demonstrate a natural explanation for this phenomenon 
and argue that the assumption of dissipationless behavior 
in superfluid systems cannot always be taken for granted.


The two-fluid model proposes that a superflow of $^4$He carries 
no thermal energy, which appears to be supported by various
 experimental observations. However, these experiments lack 
effective control of the superflow velocity, and the
 negligible thermal energy density of a superflow is 
associated with a large velocity. In Section V, we 
present an experiment in which the superflow velocity 
is partially regulated, and we demonstrate that a superflow can 
indeed carry significant thermal energy, resulting in a 
counter-intuitive heating phenomenon. This experimental result 
directly contradicts the pivotal hypothesis of the two-fluid model.

The aforementioned limitations of the two-fluid model demonstrate 
its inability to offer a comprehensive account of superfluid $^4$He. 
It is necessary to develop a full quantum microscopic description
 of the superfluid $^4$He in order to establish a reliable, 
inherent, and unified understanding of the system.

\section*{III. Grouping Behavior of Low-Lying Levels of Superfluid $^4$He}

\begin{figure}
\begin{center}
\includegraphics[scale=0.4]{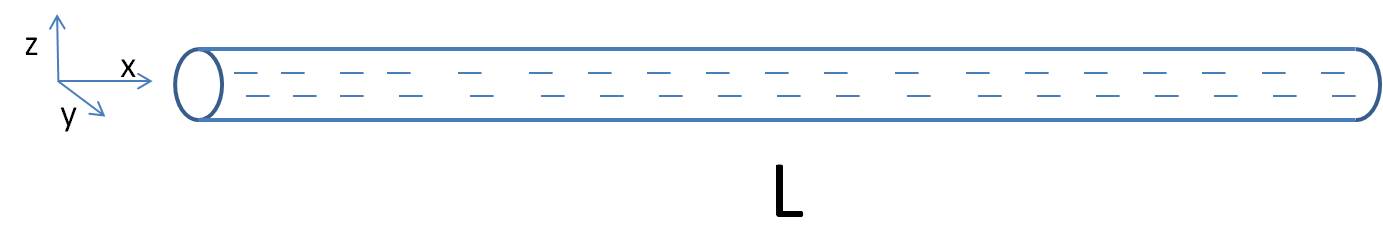}
\caption{The geometry of a liquid $^4$He system with a periodic 
length of $L$ in the $x$ direction ($x$ is identical to $x+L$).}
\label{fig:geometry}
\end{center}
\end{figure}

We consider a liquid $^4$He system that is periodic and translationally 
invariant along the $x$ axis, and is constrained in the $y,z$ directions 
by the inner boundary of a container (see Fig. \ref{fig:geometry}). The 
length of the system in the $x$ direction is $L$. The Hamiltonian of 
the system can be written as

\begin{equation}
\widehat{H} = \sum_{i=1}^{N}- \frac{\hbar^2}{2M} \nabla^2_i + \sum_{i<j}^{N}V(\mathbf{r}_i - \mathbf{r}_j),
\label{eq:Hamiltonian}
\end{equation}

where $M$ is the mass of a $^4$He atom, $\hbar$ is 
the reduced Planck constant, $N$ is the total number of atoms, and $V$ is the interaction between two atoms.

The eigen-wavefunctions of the Hamiltonian operator are governed by the following equation:

\begin{equation}
\widehat{H} \psi_\alpha (\mathbf{r}_1, \mathbf{r}_2, ..., \mathbf{r}_N)= 
E_\alpha \psi_\alpha(\mathbf{r}_1, \mathbf{r}_2, ..., \mathbf{r}_N),
\end{equation}

where $\alpha$ labels the eigen-wavefunctions and $E_\alpha$ is the eigen energy of the state.

Given that $\widehat{H}$ commutes with the total momentum operator 
$\widehat{P}_x = -\hbar \sum_{j=1}^N \frac{\partial}{\partial x_j}$ (along the $x$ axis), 
it follows that the eigen-wavefunctions of $\widehat{H}$ can be chosen to be 
the eigen-wavefunctions of $\widehat{P}_x$ simultaneously:

\begin{equation}
\widehat{P}_x \psi_\alpha (\mathbf{r}_1, \mathbf{r}_2, ..., \mathbf{r}_N)= 
{P_\alpha} \psi_\alpha (\mathbf{r}_1, \mathbf{r}_2, ..., \mathbf{r}_N),
\end{equation}

where $P_\alpha$ is the eigen momentum.

For a given eigen-wavefunction $\psi_\beta(\mathbf{r}_1, \mathbf{r}_2, ..., \mathbf{r}_N)$ with 
an energy of $E_\beta$ and a momentum of $P_\beta$, additional eigen-wavefunctions 
can be constructed by applying a Galilean transformation, which captures
 center-of-mass motion. The transformed wavefunction is obtained by,
\begin{equation}
 \psi^{cm_k}_\beta (\mathbf{r}_1, \mathbf{r}_2, ..., \mathbf{r}_N)= 
e^{i\sum_{j=1}^N  2\pi k x_j/L}  \psi_\beta (\mathbf{r}_1, \mathbf{r}_2, ..., \mathbf{r}_N),
\label{eq:trans}
\end{equation}
where $k=\pm 1,\pm 2, \pm3, ...,$ is an integer.

The energy and momentum associated with $\psi^{cm_k}_\beta$ can be determined by,
\begin{equation}
E^{cm_k}_\beta = E_\beta + (P_\beta + 2 \pi k N \hbar/ L)^2/(2NM) - P_\beta^2/(2NM)
\label{eq:transE}
\end{equation}
and
\begin{equation}
P^{cm_k}_\beta= P_\beta + 2 \pi k N \hbar/ L.
\label{eq:transP}
\end{equation}
Equations (\ref{eq:trans}), (\ref{eq:transE}), and (\ref{eq:transP}) reflect
 the Galilean invariance of the system \cite{bloch}.

We shall investigate the system's low-lying levels, which are the 
relevant levels at low temperature. Due to the exponential Boltzmann 
factor, the contribution of high-lying levels is negligible.

A grouping behavior of the low-lying levels can be revealed by
 a fundamental analysis of quantum transitions between these
 levels. These quantum transitions also play an essential role
 in the thermalization of the system. The microscopic atomic-molecular 
interactions between the system and its surroundings cause frequent 
momentum-energy exchanges and drive the quantum transitions 
between the system's low-lying levels, ultimately leading 
to thermal occupations of its levels.


Formally, the transition probability from one level with an 
eigen-wavefunction $\psi_i(\mathbf{r}_1,\mathbf{r}_2, ..., \mathbf{r}_N)$ to another 
level with a wavefunction $\psi_j(\mathbf{r}_1,\mathbf{r}_2, ..., \mathbf{r}_N)$ 
is determined through a one-particle scattering process as follows:
\begin{equation}
P_{\psi_i\to\psi_j} = |\langle \psi_j(\mathbf{r}_1, \mathbf{r}_2, ..., \mathbf{r}_N)
 | \sum_{m,l=0,\pm1, ...} f(m-l) a^\dagger_m a_l |\psi_i(\mathbf{r}_1, \mathbf{r}_2, ..., \mathbf{r}_N) \rangle|^2,
\end{equation}
where $a_l$ is the annihilation operator corresponding to a single-particle 
orbit that carries an eigenmomentum of ${2\pi l\hbar/L}$ along the $x$-direction 
(the wavefunction of the orbit has an $x$-dependence of $e^{i 2\pi l x/L}$, and the
 $y,z$-dependence of the orbit is ignored for simplicity). $a^\dagger_m$ is the
 creation operator for a single-particle orbit with a wavefunction of $e^{i 2\pi m x/L}$, 
and the function $f(m-l)$ describes the strength of the scattering.

Depending on these two levels, $P_{\psi_i\to\psi_j}$ can easily vanish. 
In some instances it may assume an exact value of zero, whereas 
in the realistic case, it vanishes in the sense that its value
 decreased exponentially as a function of $N$.  
Whether or not $P_{\psi_i\to\psi_j}$ vanishes is dependent on 
 the degree of proximity between  the two corresponding wavefunctions. 
For instance, let $\psi_i $ 
have a Bose-Einstein condensation (BEC) form
which is associated with a single-particle orbit $|\phi_1\rangle $ with  
a wavefunction of $\phi(y,z) e^{i 2 \pi x/L} $, such that 
$\psi_i(\mathbf{r}_1,\mathbf{r}_2, ..., \mathbf{r}_N )= |\phi_1^N\rangle $.
Similarly, let $\psi_j $ represent the BEC of another single-particle orbit
 $|\phi_2\rangle =\phi(y,z) e^{i 4 \pi x/L }$, resulting in 
$\psi_j(\mathbf{r}_1, \mathbf{r}_2, ..., \mathbf{r}_N)= |\phi_2^N \rangle $.
it can be shown easily that  
$ P_{\psi_i\to\psi_j}$  is exactly zero.
Furthemore, consider $\psi_i(\mathbf{r}_1, \mathbf{r}_2, ..., \mathbf{r}_N)= |\phi_1^{M_1} \phi_2^{N-M_1}\rangle$ ( 
where $M_1$ particles 
occupy orbit $\phi_1$ and $N-M_1$ particles occupy orbit $\phi_2$, $M_1 \leq N$),   and 
  that $ \psi_j(\mathbf{r}_1, \mathbf{r}_2, ..., \mathbf{r}_N)= |\phi_1^ {M_2}\phi_2^{N-M_2}\rangle$ ,
$P_{\psi_i\to\psi_j}$ always vanishes unless $|M_1- M_2| \leq 1$. 
Here, the degree of difference between 
  $|\phi_1^{M_1} \phi_2^{N-M_1}\rangle$
and $|\phi_1^ {M_2}\phi_2^{N-M_2}\rangle$ can be roughly approximated as $|M_1- M_2|$, 
which is the difference of 
number of particles occupying a major orbit. 
These straightforward observations suggest that 
 the transition probability between two quantum states associated 
with partial BEC
  is negligible, unless
 they have almost an equal number of particles to occupy 
the same orbit.
  
We shall consider the ground state wavefunction of the 
system $\psi_g(\mathbf{r}_1, \mathbf{r}_2, ..., \mathbf{r}_N)$, which has 
the ground state energy $E_g$ ($\widehat{H}|\psi_g\rangle=E_g|\psi_g\rangle$) and 
has zero momentum along the $x$ axis ($P_x|\psi_g\rangle=0$). One can obtain 
another eigenstate which is a Galilean transformation of the ground state: 
$\psi_g^{cm_1}(\mathbf{r}_1, \mathbf{r}_2, ..., \mathbf{r}_N)=
e^{i\sum_{j=1}^N2\pi x_j/L}\psi_g(\mathbf{r}_1, \mathbf{r}_2, ..., \mathbf{r}_N)$. 
The transition probability between these two states is given by
$P_{\psi_g\to\psi_g^{cm_1}}=\left|\langle\psi_g^{cm_1}|\sum_{m,l}f(m-l)a_m^\dagger a_l|\psi_g\rangle\right|^2$.
It can be argued that $P_{\psi_g\to\psi_g^{cm_1}}$ vanishes naturally. First, 
the momentum difference between these states is $2\pi N\hbar/L$, which 
is large from a microscopic perspective. For such a transition to occur, it would 
require a scattering process involving single-particle orbits with large momentum. 
However, for the ground state, which can be written as a superposition of 
Fock states in momentum space, its component involving a single-particle 
orbit of large momentum is negligible, as the engagement of a large 
momentum (single-particle) orbit costs high kinetic energy, which is 
unfavorable to the ground state. As a result, the scattering amplitude
 vanishes. Secondly, the ground state is expected to exhibit strong correlation
 of $^4$He atoms. It is difficult to precisely describe this correlation, but 
a substantial portion of it is manifested by a partial Bose-Einstein 
condensate (BEC) in the wavefunction \cite{london1,london2,tisza}. The condensate 
fraction of the ground state is estimated to be around $10\%$ in the 
literature \cite{BECfrac_Puff,BECfrac_Harling,BECfrac_Rodriquez,BECfrac_Svensson,
BECfrac_Svensson2,BECfrac_Sears,BECfrac_Ceperley,BECfrac_Whitlock,
BECfrac_Sosnick,BECfrac_Glyde,BECfrac_Boninsegni,BECfrac_Boninsegni2,
BECfrac_Rota,BECfrac_Prisk,BECfrac_Mahan,BECfrac_Glyde2}. For convenience, 
if a many-body eigenstate involves a partial BEC, the single-particle 
orbit required to accommodate this BEC is referred to as the base 
orbit of the state. It is apparent that the base orbit of $\psi_g$ possesses 
zero momentum along the $x$-direction. On the other hand, $\psi_g^{cm_1}$ 
corresponds to a Galilean transformation of $\psi_g$, while the internal 
correlations of the atoms are unaffected by the transformation. However, 
the base orbit of $\psi_g^{cm_1}$ has an eigen momentum 
value of $2\pi\hbar/L$, and it is orthogonal to the base orbit of $\psi_g$. 
According to the previous discussions, there is a large degree of difference
 between $\psi_g$ and $\psi_g^{cm_1}$; therefore, $P_{\psi_g\to\psi_g^{cm_1}}$ shall vanish.


Not only the ground state, but also all low-lying eigenstates possess the 
same type of correlation, which is manifested in a crude form of a 
partial Bose-Einstein condensate (BEC) \cite{BECfrac_Svensson2, BECfrac_Sears, 
BECfrac_Sosnick, BECfrac_Glyde, BECfrac_Boninsegni2, BECfrac_Rota, 
BECfrac_Prisk, BECfrac_Glyde2}. This correlation is primarily due to 
the effect of Bose exchange interaction \cite{yu1}. If a quantum 
state involves a condensate fraction, the energy of the Bose
 exchange interaction can be significantly reduced. In contrast, 
for an eigenstate without a condensate fraction, the exchange 
interaction substantially raises its energy, excluding it from 
the low energy regime.



Following the preceding analysis, a grouping behavior for 
low-lying levels arises naturally. Specifically, the levels 
sharing base orbit are categorized together, forming a
group separate from other groups comprised of
levels with  different base orbits.

 The group involving the ground state $\psi_g$ is denoted 
as $grp_0$, while the group incorporating $\psi_g^{cm_1}$ is
 labeled as $grp_1$. It is evident that $grp_1$ can be viewed
 as a Galilean transformation of $grp_0$. Likewise, $grp_k$ ($k=2, 3, \dots$)
 can be obtained by multiplying all eigen wavefunctions in $grp_0$ by 
a Galilean factor of $e^{i\sum_{j=1}^N 2\pi k x_j/L}$. The group 
$grp_{-|k|}$ with a negative integer value of $k$ can be 
formed by applying a corresponding Galilean factor.

For any two levels ($\psi_a$ and $\psi_b$) in the
 same group, it is always possible to find a chain of 
levels ($\psi_n,n=1,2,...,N^c)$ in this group such that all the transition 
probabilities $P_{\psi^a\to\psi_1}, P_{\psi_1\to\psi_2}, 
P_{\psi_2\to\psi_3},..., P_{\psi_{N^c-1}\to\psi_{N^c}},
 P_{\psi_{N^c}\to\psi_b}$ do not vanish. In other words, 
$\psi_a$ can eventually be scattered to $\psi_b$ through a 
series of scattering processes. However, if two levels belong 
to two different groups, it is not possible to find a
 series of scattering processes that connects these two levels 
unless some high-lying levels are involved.

At low temperatures, the system can establish a thermal 
equilibrium with its surroundings that is specific to each group. 
For instance, if some levels of a particular group are initially 
occupied, the frequent quantum exchanges between the system and 
its surroundings will lead to the dispersion of the level 
occupations, eventually resulting in a thermal population of 
all levels in the group. On the other hand, 
 groups that are initially unoccupied remain vacant as 
inter-group transitions are prevented by high energy barriers.

\begin{figure}
\begin{center}
\includegraphics[scale=0.65, angle=0]{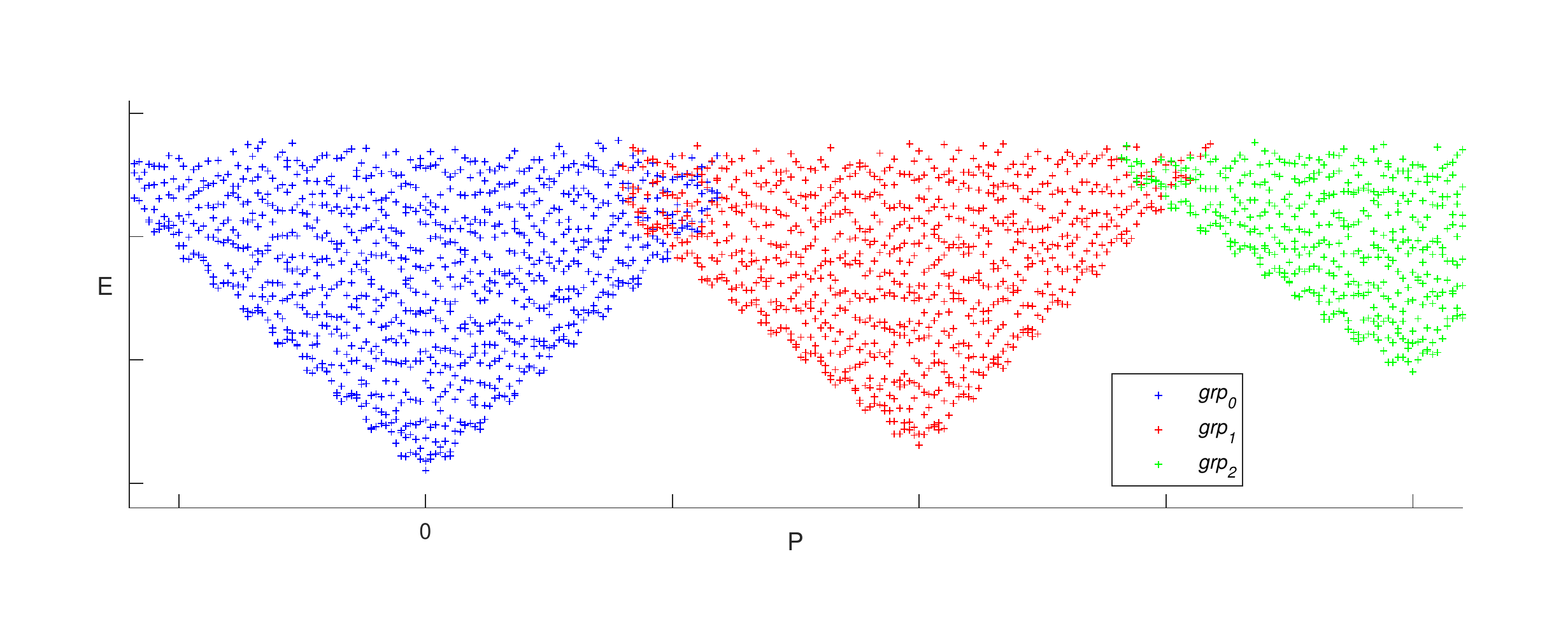}
\caption{A schematic plot of some low-lying levels of $grp_0$, $grp_1$, 
and $grp_2$, labeled in color, in the momentum-energy plane. Each
 level is marked by a plus sign. The groups overlap in certain regions
 of the plane, and the extent of their overlap increases in regions of high energy
 (not depicted in the figure).}
\label{fig:grp123}
\end{center}
\end{figure}

Some levels in $grp_{0}$, $grp_{1}$, and $grp_{2}$ 
are schematically plotted in the momentum-energy plane in Fig. \ref{fig:grp123}. 
The lower boundaries (the many-body dispersion lines) of 
$grp_{0}$ are approximately linear in both positive and 
negative momentum directions, and the slopes of the two linear 
dispersions are of equal magnitude \cite{yu1, yu2, 
bogoliubov, bloch}. In $grp_{1}$, the magnitude of the 
positive linear dispersion slope is larger than that of 
the negative linear branch slope.   
 This is because the Galilean 
transformation of a positive linear dispersion line 
(in the positive $P$ direction) makes the line more tilted, whereas
 it makes the line less tilted in the case of negative linear
 dispersion. In the case of $grp_{2}$, the magnitude of 
the positive linear dispersion line increases further. 
Note that $grp_{0}$ and $grp_{1}$ overlap in some regions, 
with the overlap being more prevalent in the upper 
region of higher energy. The overlap of these two groups
 might give rise to a false impression that quantum jumps between 
them are possible. As analyzed earlier, every wavefunction in 
$grp_{0}$ has a large degree of difference from the wavefunctions 
in $grp_{1}$ due to their distinct base orbits. Consequently, the 
transition probability between them vanishes.


The phenomenon of superfluidity can be naturally explained 
based on the grouping of the levels. If only $grp_0$ is 
initially occupied, the average momentum of the system at 
thermal equilibrium is zero, indicating a static state. 
Assuming, for instance, that $grp_2$ is exclusively 
occupied at the outset, the average momentum of the 
system at thermal equilibrium can be determined 
statistically by the following equation:
\begin{equation}
\overline{P}^{T}_{grp_2} = \frac{\sum_{ \phi_{\gamma} \in grp_2}
\langle \phi_{\gamma} |\widehat {P_x} |\phi_{\gamma}\rangle e^{-\frac{E_\gamma}{kT}}}
{\sum_{ \phi_{\gamma} \in grp_2}
\langle \phi_{\gamma} | \phi_{\gamma}\rangle e^{-\frac{E_\gamma}{kT}}}.
\label{eq:momentumP}
\end{equation}
A non-zero value of $\overline{P}^{T}_{grp_2}$ indicates 
the presence of a persistent current in the system. $\overline{P}^{T}_{grp_2}$ 
possesses a temperature dependence due to the Boltzmann factor 
in Eq. \ref{eq:momentumP}. 
If the temperature of the system undergoes a gradual change followed by 
a return to its initial value, $\overline{P}^{T}_{grp_2}$ will vary 
accordingly and eventually revert to its initial value. Such a temperature 
dependence is largely confirmed by an experiment in the past \cite{reppy}.


In reality, the translational invariance of a superfluid $^4$He 
system is imperfect. For example, irregularities in the inner 
wall of the container can disrupt exact translational symmetry. 
Nonetheless, this symmetry-broken interaction or potential can 
be considered a minor perturbation term when compared to 
the terms in Eq. \ref{eq:Hamiltonian}. The atomic-molecular 
interactions between the system and the container are 
confined to a few layers of $^4$He atoms near the wall, and 
the interaction strength decreases quickly as a $^4$He atom 
moves away from the wall surface.

Although the microscopic many-body wavefunctions are no 
longer the eigenstates of $\widehat{P_x}$, one can use 
the expectation value of this operator for theoretical 
purposes. In higher order treatments, this perturbation 
term might mix the levels of the same group by an 
infinitesimal amount, but it is not sufficient to mix two 
levels belonging to two distinct groups. Inter-group mixing 
can only occur if the perturbation term has a homogeneous 
influence on all atoms in the system. Therefore, it can be
 concluded that the grouping behavior of microscopic levels remains 
quite robust, regardless of the exact translational symmetry.

\begin{figure}
\begin{center}
\includegraphics[scale=0.4, angle=0]{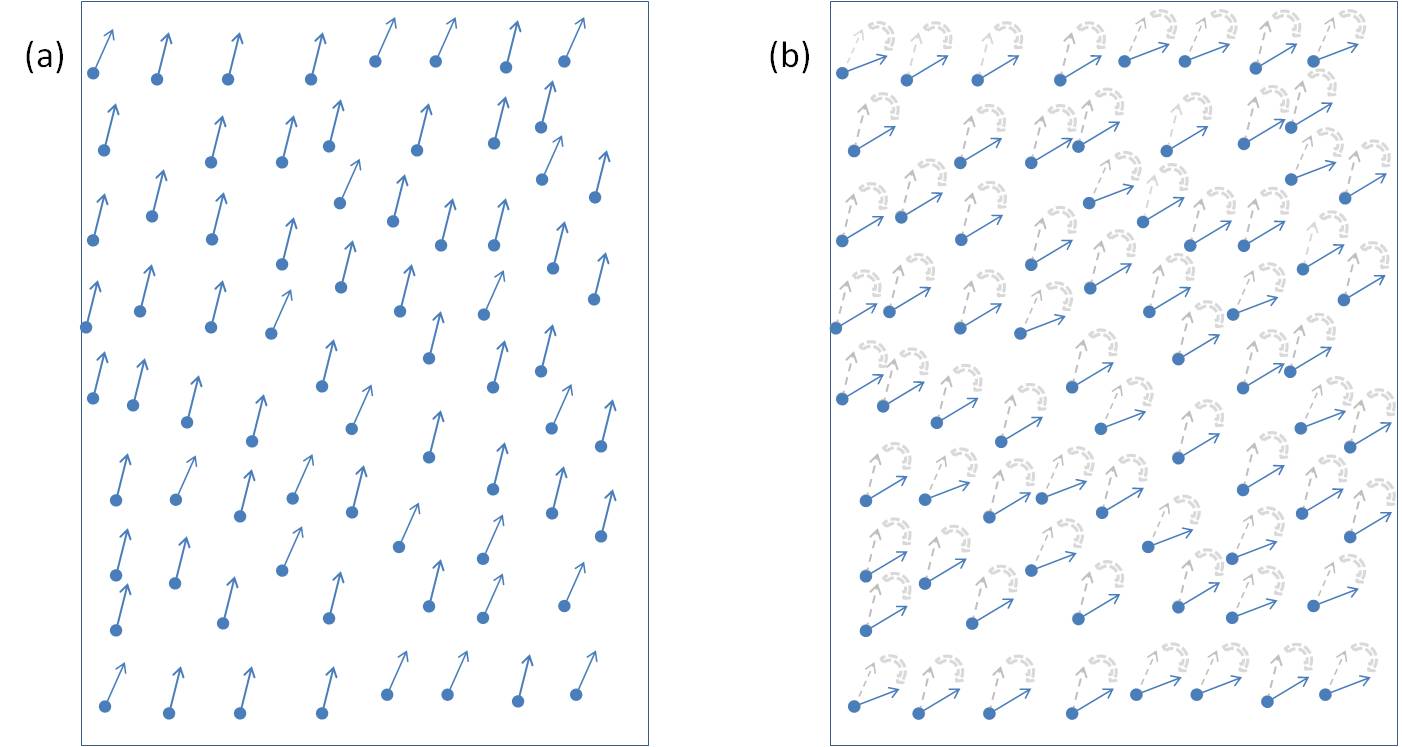}
\caption{a) In a ferromagnet, the microscopic magnetic 
moments (spins) roughly align in
 the same direction. b) A smooth transition between 
two metastable states is
 only possible if all the microscopic moments rotate 
collectively, which 
generally requires external manipulation.} %
\label{fig:ferro}%
\end{center}
\end{figure}

The grouping behavior of low-lying levels in superfluid $^4$He is 
of a fully quantum nature. It is related to a particular property
 of each low-lying level, namely, the corresponding wavefunction 
has a condensate fraction, as well as transition probabilities 
among these levels. Although it was unnoticed in the past, one 
can realize that the grouping of microscopic levels is not 
unique to a superfluid system. For instance, consider a ferromagnet
 as an illustration. Below the Curie point, the microscopic magnetic 
moments (spins) in the ferromagnet tend to align approximately in 
the same direction (see Fig. \ref{fig:ferro}(a)). The system possesses 
numerous macroscopic states due to the different directions of total 
magnetic moments, and each of these states corresponds to a specific 
group of microscopic levels. The metastability of each macroscopic 
state can be attributed to the following factors: (i) the thermal 
transition between any two distinct groups of microscopic levels 
is impossible at low temperatures since such a transition 
requires overcoming a high energy barrier that separates the 
groups, and (ii) a smooth transition between two different groups 
of levels is possible if all spins rotate simultaneously by the 
same angle (see Fig. \ref{fig:ferro}(b)). This collective motion 
of all spins involved in such a transition does not encounter a 
high energy barrier. However, this collective motion cannot 
initiate spontaneously and requires some form of external operation, such 
as the application of a strong magnetic field. It should be noted that, 
in the case of superfluid $^4$He, a smooth transition between different 
groups of levels can occur under the influence of a global force. For
 instance, the application of a force, such as gravity or hydrodynamic 
pressure, that acts uniformly on all $^4$He atoms and accelerates 
them collectively can establish such a transition.

\begin{figure}
\begin{center}
\includegraphics[width=1.1\textwidth]{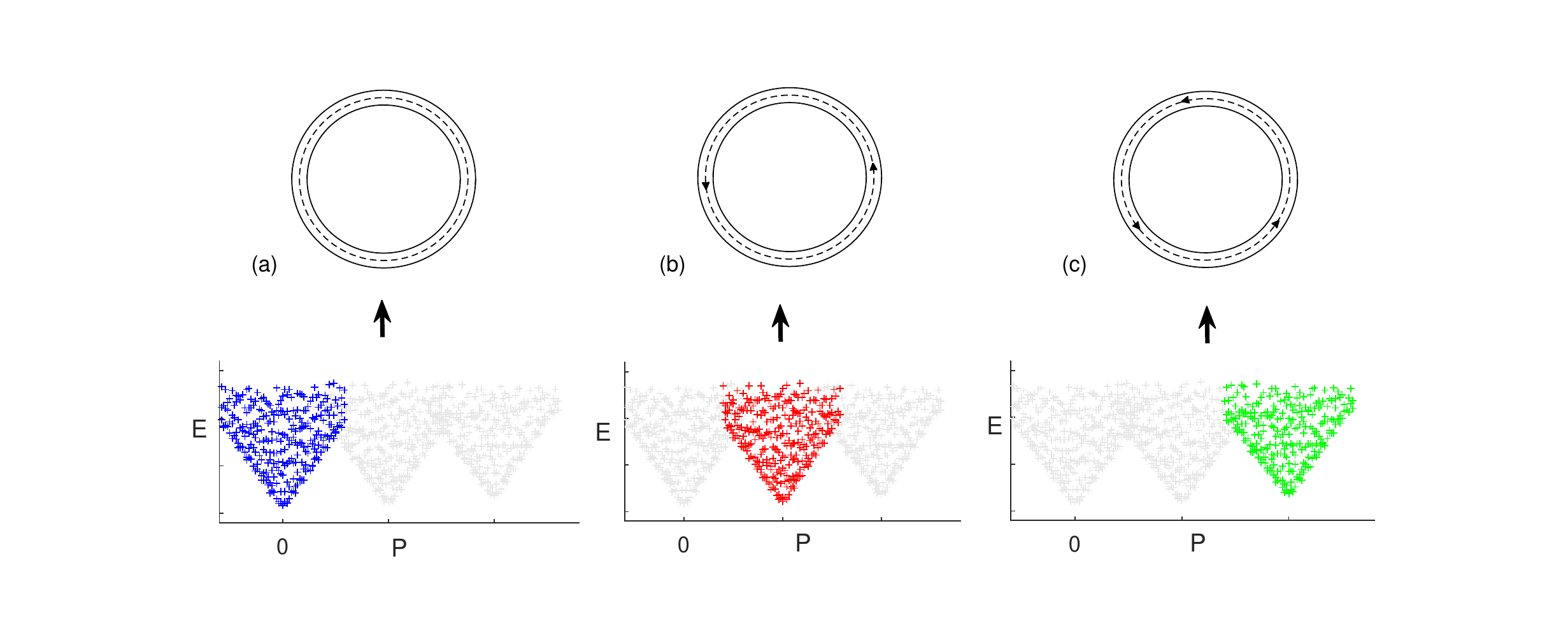}
\caption{A schematic plot of macroscopic-microscopic correspondence 
that underlies the metastability of $^4$He superflows. (a) represents a 
static state, while (b) and (c) show superflow states with small and 
slightly larger velocities, respectively. The occupied groups 
are highlighted in color.} %
\label{fig:metaphase}%
\end{center}
\end{figure}

In the context of the analogy between the superfluid $^4$He and 
a ferromagnet, a broader perspective on the grouping of the microscopic
 can be formed. In such systems, two fundamental facts are observed: 
i) the existence of a number of macroscopic metastable states at 
low temperatures; and ii) a large number of microscopic 
low-lying levels. Establishing a connection between these two 
facts shall naturally lead to the realization that the 
low-lying levels of the system are organized into groups, 
physically separated by some high energy barriers. In 
this way, a macroscopic metastable state corresponds 
to a group of microscopic levels (such a macroscopic-microscopic 
correspondence resolves the mystery of superfluidity in 
a transparent way, see Fig. \ref{fig:metaphase}). On the other 
hand, at sufficiently high temperatures, the energy 
barriers are overcome and all groups of low-lying levels are 
thermally occupied along with the relevant high-lying levels, 
resulting in a reduction of the number of macroscopic 
thermal states to one.

The preceding discussions can be readily generalized to other condensed 
matter systems that exhibit low-temperature phases characterized by an order 
parameter. The microscopic quantum pictures of these systems share the following features:

i) There exists a grouping behavior of the low-lying levels in each system. The 
occupancy of a specific group and the vacancy of other groups correspond to 
a metastable state with a certain order parameter value that 
is statistically determined by the occupied group.

ii) A smooth inter-group transition at low temperature might be 
feasible if a suitable collective motion of all particles in 
the system occurs. Still, such collective motion is not guaranteed. 
Inter-group transitions through spontaneous thermal excitations 
are prohibited since they involve high-lying levels that are 
effectively irrelevant at low temperatures.

iii) As the temperature rises above the transition point, the 
grouping of low-lying levels becomes irrelevant, and inter-group 
transitions become possible through thermal excitations involving 
high-lying levels, which are more abundant and statistically more 
significant than the low-lying levels. This leads the system to 
a single thermal state with an order parameter value of zero.

We shall explain why superfluid $^4$He is not dissipationless
 in certain cases (some relevant experiments were discussed in Section II). 
The dissipationless behavior of superfluid $^4$He is predicated on 
its attainment of group-specific thermal equilibrium, which 
is manifested by a steady flow motion. However, when the superflow 
oscillates between two vessels, there is frequent microscopic 
inter-group transitions caused by a global force which drives 
all $^4$He atoms in the flow. As a result, numerous groups of 
low-lying levels become intermittently occupied and unoccupied 
over time, which precludes the superflow from reaching thermal 
equilibrium. Inevitably, dissipation arises during these 
microscopic transition processes, and the system behaves similarly 
to a normal system regarding the temperature-dependent 
dissipation rate at low temperature limits.

\begin{figure}
\centering
\includegraphics[scale=0.33]{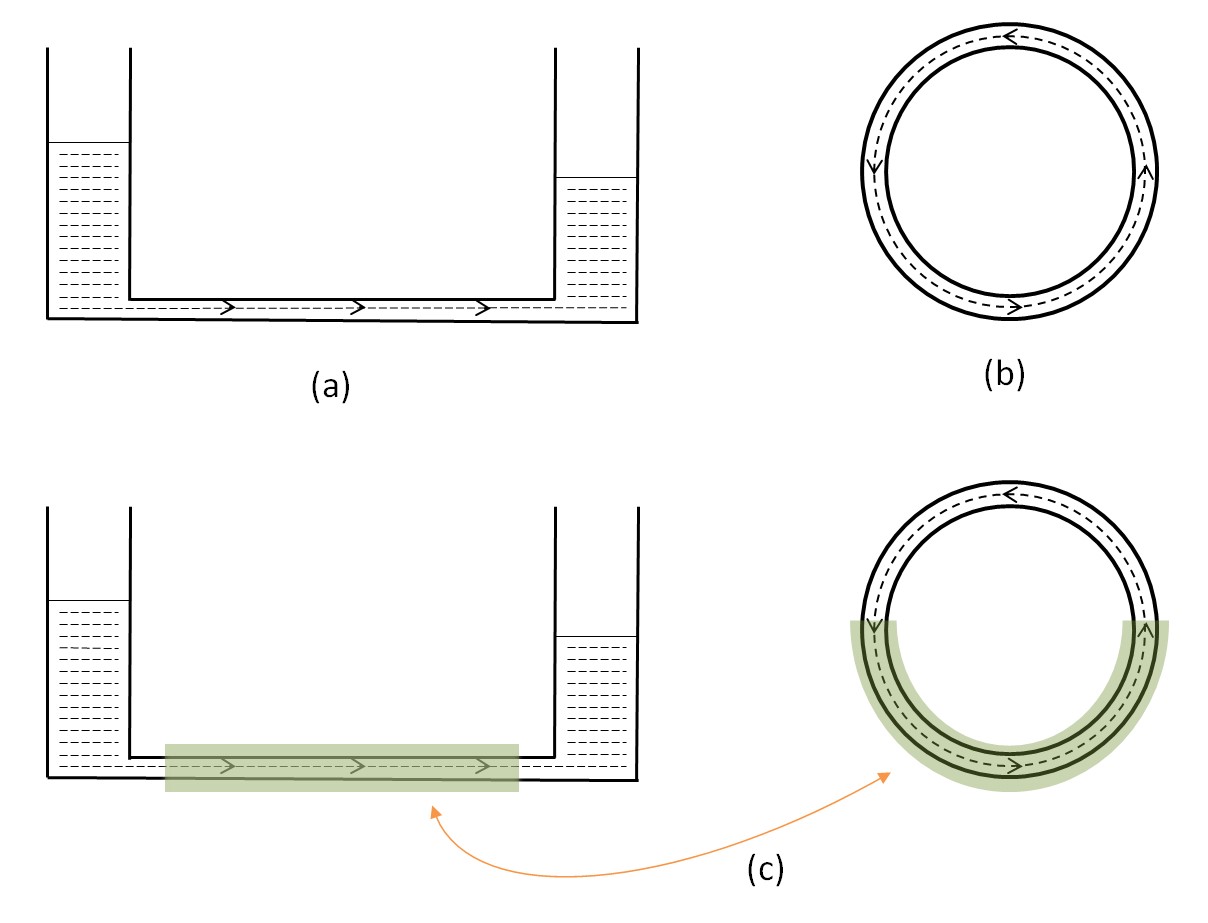}
\caption{(a) An open superflow system of $^4$He which connects to two vessels. 
(b) A closed superflow system with a naturally periodic boundary condition. 
(c) The systems within the colored shaded areas are equivalent to each other.}
\label{fig:boundaries}
\end{figure}

In order to facilitate a thorough understanding of the fundamental 
properties of superfluid $^4$He, it is useful to reflect on one of the 
theoretical conditions involved in the discussions. It has been shown 
that the low-lying energy levels of a superfluid $^4$He system exhibit 
a characteristic grouping behavior. One might naturally wonder 
whether this behavior is dependent solely on the periodic boundary 
condition of the system under analysis. However, further examination 
reveals that this is not necessarily the case.

In reality, it is generally unreasonable to assume 
that an intrinsic property of a physical system is dependent 
upon its boundary conditions. In the context of $^4$He superflow,
 a closed system naturally possesses a periodic boundary 
condition, whereas an open system can be deemed a part 
of a closed system (see Fig. \ref{fig:boundaries}), 
indicating that both systems possess the same fundamental properties.

Upon closer examination, one can realize that the grouping of
low-lying energy levels in superfluid $^4$He is attributed to 
two fundamental physical factors: Galilean invariance and the 
discreteness of single-particle orbits carrying momentum 
along the superflow direction, which arises from the finite 
dimensions of the system \cite{finitesize}. As both open and \
closed systems possess these two factors, they 
intrinsically possess the same grouping property. While the 
periodic boundary condition of a closed system provides a 
more straightforward means of demonstrating this property 
from a formal standpoint, the boundary condition of an 
open system may introduce additional formal complexity.

\section* {VI.  Velocity dependence of the thermal energy of 
a superflow}

The preceding section expounded on the grouping behavior 
of microscopic levels of superfluid $^4$He. It is evident 
that the system's macroscopic properties are statistically 
determined by the occupied group(s) and naturally have a 
dependence on the group(s). In this regard, we shall 
examine two important properties: the superflow velocity 
and the thermal energy (density).
\begin{figure}
\begin{center}
\includegraphics[scale=0.8]{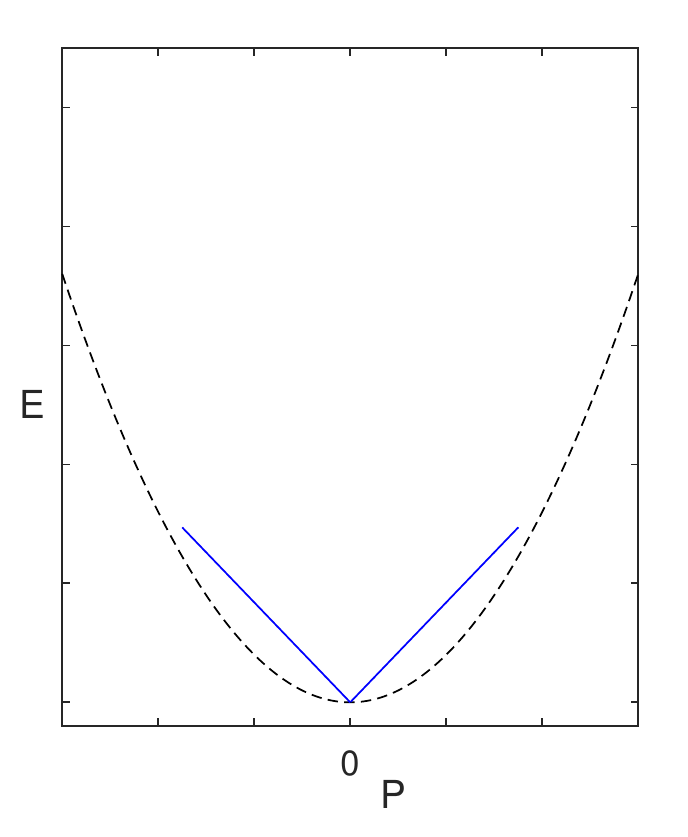}
\caption{The blue lines in the figure represent 
the linear dispersions of $grp_0$, while the dashed 
parabola corresponds to the energy-momentum relation
 $E(P)= E_g + P^2/2NM$. Since all energy levels of 
the system must be located above the parabola, the 
blue lines cannot extend further to cross the curve. As 
a result, the dispersion behavior must be adjusted at 
large $|P|$ regions so that the extended dispersion 
lines can remain above the parabola, as shown in
 Fig. \ref{fig:lineardispersion}(a).}
\label{fig:parabolic}
\end{center}
\end{figure}

To further characterize the group $grp_k$, a velocity parameter
$v_k$ can be defined as:
\begin{equation}
v_k = \frac{2\pi \hbar k}{M L}.
\end{equation}
This parameter corresponds to the center-of-mass velocity
of the microscopic state with a wavefunction of
$e^{\sum_{j=1}^N i 2\pi k x_j/L} \psi_g$, which
belongs to the group $grp_k$.

The thermally averaged velocity of the system in the group
$grp_k$ is determined by:
\begin{equation}
\overline{v}^{T}_{grp_k} = \frac{\sum_ { \phi_{\gamma} \in grp_k}
\langle \phi_{\gamma} |\frac{\widehat {P_x}}{N M} |\phi_{\gamma}\rangle e^{\frac{-E_\gamma}{kT}}}
{\sum_{ \phi{\gamma} \in grp_k}
\langle \phi_{\gamma} | \phi_{\gamma}\rangle e^{\frac{-E_\gamma}{kT}}}.
\end{equation}

The superflow velocity is clearly represented by
$\overline{v}^{T}_{grp_k}$, which possesses a temperature 
dependence and can differ significantly from $v_k$. Empirical
evidence from superfluid $^4$He indicates
that $v_k$ can reach velocities several tens of meters per second or more, while
the corresponding $\overline{v}^{T}_{grp_k}$ at temperatures well
below the transition point
is generally two orders of magnitude smaller.
Moreover, $\overline{v}^{T}_{grp_k}$ approaches zero
near the transition point, whereas $v_k$ remains independent of temperature.
At a specific temperature (above $1$ $K$),
$\overline{v}^{T}_{grp_k}$ increases with the group number $k$ and
may be used to distinguish the groups. Consequently, other group-specific
properties of the system can be regarded as being flow-velocity dependent.
We shall investigate the flow-velocity dependence of the
superflow's thermal energy.

\begin{figure}
\centering
\includegraphics[width=1.1\textwidth]{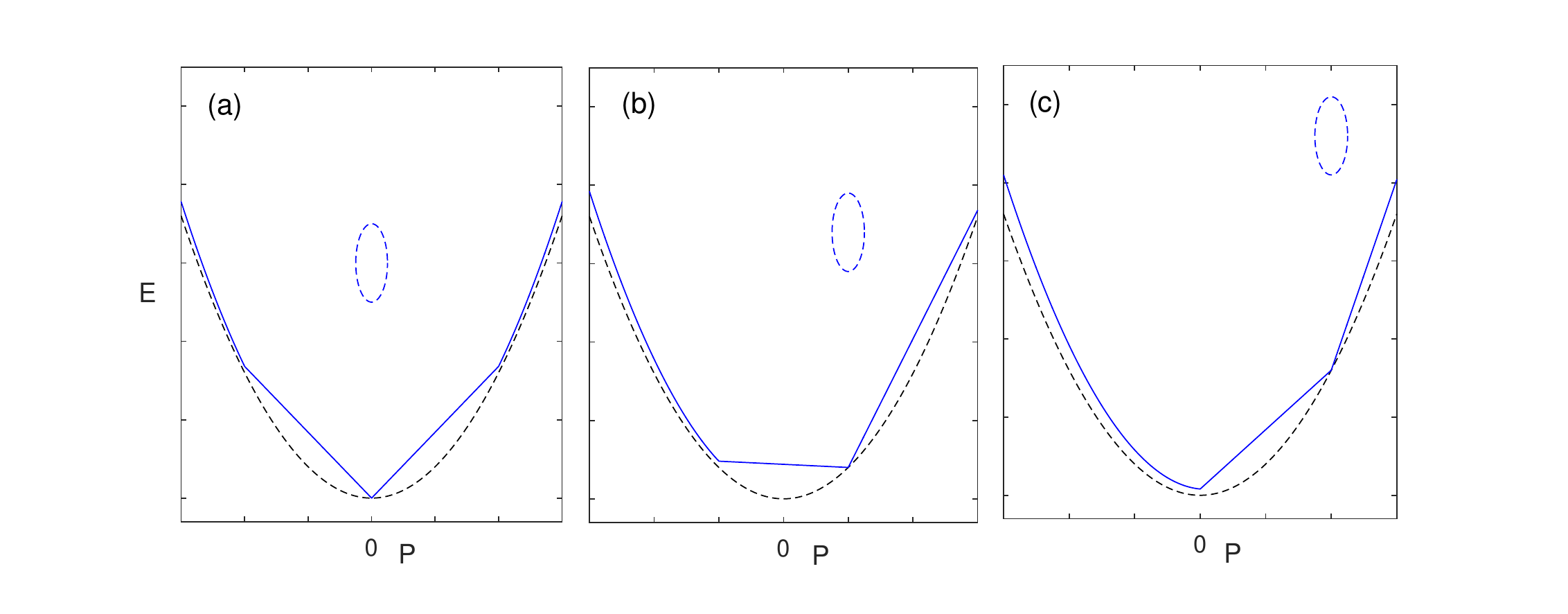}
\caption{(a) The magnitude of the slope of the linear dispersion 
line of $grp_0$ remains constant. The dashed parabola corresponds 
to the energy-momentum relation $E(P) = E_g + P^2/2NM$.  (b) and (c) 
show two Galilean transformations of $grp_0$, corresponding to 
superflow states with middle and large flow velocities, respectively. 
The ellipse in (a) roughly denotes the region of dense energy 
levels that make the major contribution to the system's thermal 
energy at a given temperature. In (b) and (c), the ellipse 
region is raised, and the dense levels in this region are 
less involved in the contribution to the thermal energy, 
resulting in a decrease in the thermal energy of the system.}
\label{fig:lineardispersion}
\end{figure}

\begin{figure}
\centering
\includegraphics[width=1.1\textwidth]{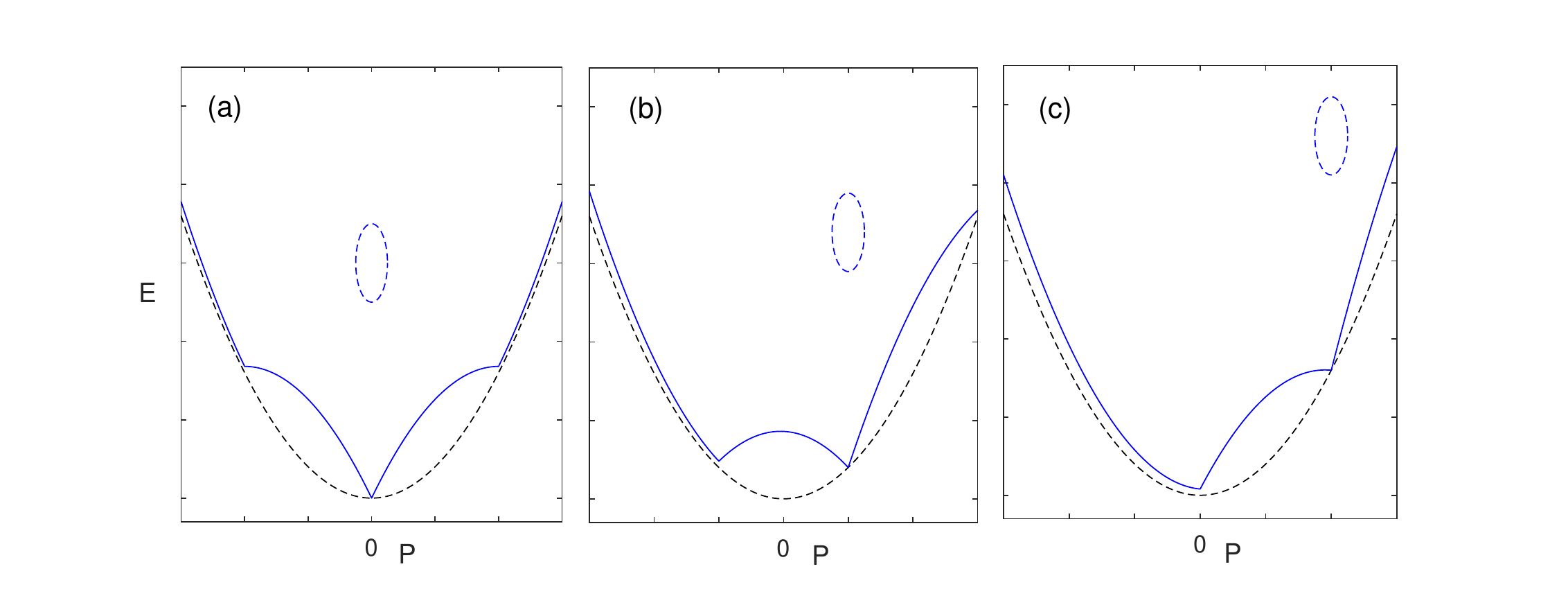}
\caption{The dispersion of $grp_0$ in this figure differs from 
that in Fig. \ref{fig:lineardispersion}(a). At small momentum 
values, the dispersion is linear, although the magnitude of 
the slope decreases as $|P|$ increases. (b) and (c) display 
two Galilean transformations of $grp_0$.}
\label{fig:sub_lineardispersion}
\end{figure}

The thermal energy of the system in the group $grp_k$ is 
formally calculated as follows:
\begin{equation}
\overline{E}^{T}_{grp_k} = \frac{\sum_{\phi_{\gamma} 
\in grp_k} \langle \phi_{\gamma} |E_\gamma |\phi_{\gamma}\rangle e^{\frac{-E_\gamma}{kT}}}
{\sum_{\phi_{\gamma} \in grp_k} \langle \phi_{\gamma} 
|\phi_{\gamma}\rangle e^{\frac{-E_\gamma}{kT}}} .
\label{eq:thermalE}
\end{equation}
In order to perform a qualitative analysis of $\overline{E}^{T}_{grp_k}$,
it is imperative to possess a comprehensive understanding of
the level distributions of $grp_k$ in the momentum-energy plane. One can
acquire this knowledge by investigating solely the level distributions of $grp_0$, as
$grp_k$ can be considered a Galilean transformation of $grp_0$. 
In Fig. \ref{fig:grp123}, only a subset of
levels within $grp_0$ is schematically plotted, where 
the lower boundaries are presented by two
approximately linear dispersions. However, this linearity breaks down
at high momentum regimes for a fundamental reason.
In Fig. \ref{fig:parabolic}, the linear boundaries of 
$grp_0$ are plotted alongside a parabolic curve,
defined as $E = E_g + P^2/2NM$. It can be
realized that no microscopic levels of the system can 
exist below this parabolic curve. This can be proven
using the method of proof by contradiction. Suppose there 
exists an eigenstate with a wavefunction $\psi_\zeta$, which 
has a momentum of $P_\zeta$ and has an energy 
$E_\zeta < E_g + P_\zeta^2/2NM$. Then, one can find an integer 
$n$ such that $n \leq -\frac{P_\zeta L}{2 \pi \hbar} < n+1$.
Next, consider the state described by the wavefunction 
$\psi^{cm_n}_\zeta = e^{\sum_{j=1}^N i 2\pi n x_j/L} \psi_\zeta$. This 
state has a momentum close to zero and an energy of 
approximately $E^{cm_n}_\zeta = E_\zeta - P_\zeta^2/2NM$, which 
is smaller than $E_g$. However, this contradicts the fact 
that the lowest eigen energy should be $E_g$.

In Fig. \ref{fig:lineardispersion}(a), a possible scenario 
for the dispersion behavior of ${grp_0}$ is presented, where 
the magnitude of the slopes of the linear dispersion remains 
constant until it approaches the parabola. On the other 
hand, Fig. \ref{fig:sub_lineardispersion}(a) illustrates 
another possible scenario where the magnitude of the 
slopes of the linear dispersion decreases with $|P|$ 
until it approaches the parabola.

After qualitatively illustrating the behavior of the lower 
boundaries of ${grp_0}$ in the momentum-energy plane, it 
is imperative to analyze the level density of ${grp_0}$ 
at various locations in the same plane. Notably, the 
region proximate to the group's lower boundary exhibits 
a relatively sparse distribution of levels. In contrast, 
levels are considerably more concentrated in the 
central region around $P \approx 0$, particularly 
at higher energies (with the extent of the high-energy 
region dependent on temperature). When considering the 
Boltzmann factor $e^{-E/kT}$, it is evident that the 
levels enclosed by the ellipse in Fig. 
\ref{fig:lineardispersion}(a) (and Fig. 
\ref{fig:sub_lineardispersion}(a)) primarily contribute 
to the system's thermal energy. Conversely, the contribution 
from the scarce levels in the remaining region is 
comparatively insignificant.

\begin{figure}
\begin{center}
\includegraphics[width=0.6\textwidth]{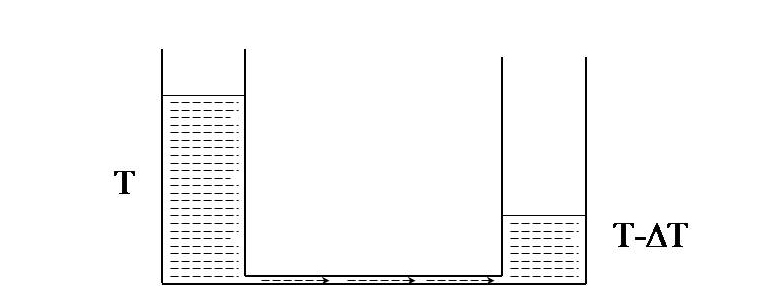}
\caption{The mechano-caloric effect of superfluid $^4$He. The 
superflow exhibits a high flow velocity and low thermal 
energy density in the narrow channel. As the superflow 
exits the channel, it transitions to a state of 
vanishing flow velocity and a correspondingly 
high thermal energy density, assuming that the 
temperature remains constant. The principle of 
energy conservation requires a decrease in temperature 
to compensate for the energy change due to the 
decrease in flow velocity.

}
\label{fig:mechanocaloric}
\end{center}
\end{figure}

When ${grp_0}$ is transformed into ${grp_k}$ 
(as plotted in Figs. \ref{fig:lineardispersion} 
and \ref{fig:sub_lineardispersion}), it becomes apparent 
that the region of high level-density is displaced 
towards the right ($k>0$) and is elevated by an energy 
quantity of approximately $\Delta E= \frac{1}{2} N M v_k^2$, as 
indicated by Eq. \ref{eq:transE}. Concurrently, the sparse regions 
located near the left boundary of ${grp_0}$ are shifted towards 
lower energies. The upliftment of the high level-density region 
leads to its reduced contribution to the thermal energy as 
a result of the Boltzmann factor in Eq. \ref{eq:thermalE}, 
thereby resulting in a decrease in the value of $\overline{E}^{T}_{grp_k}$ due 
to the diminishing major part. $\overline{E}^{T}_{grp_k}$ is a decreasing 
function of $|v_k|$ or $|k|$ owing to the upliftment of the 
high level-density region.

The energy difference $\Delta E$ is $1.3 \times 10^{-24}$ J 
per atom at $v_k = 20 m/s$, and $12 \times 10^{-24}$ J per 
atom at $v_k = 60 m/s$. At $T=1.6 K$, the average thermal 
energy of a static superfluid $^4$He is roughly 
$2.6 \times 10^{-24}$ J per atom, while at $T=1.8 K$, it 
is about $5.6 \times 10^{-24}$ J per atom. A comparison 
of these values suggests that the contribution of 
the high level-density region to the thermal density 
becomes effectively negligible when $v_k$ is sufficiently 
large. The thermal energy density of a superflow can 
decrease by several orders of magnitude as the 
flow velocity increases, which is a remarkable quantum 
property of superfluid $^4$He.

The mechano-caloric effect of superfluid $^4$He (see Fig. \ref{fig:mechanocaloric}) 
can be naturally explained in light of the velocity dependence 
of the thermal energy. Notably, when there is a substantial 
variation in the flow velocity of a superflow, the associated 
thermal energy experiences significant perturbation. 
The fundamental principle of energy conservation 
necessitates a corresponding adjustment in the 
temperature of the superflow to compensate for 
the energy variation caused by the flow velocity 
change. This intrinsic coupling between the variations 
in flow velocity and temperature reflects the 
group-specific nature of the system's thermal states.

\section*{V. Experimental observations of an 
self-heating effect of $^4$He superflows}

In this section, we present experimental observations of a
self-heating phenomenon of $^4$He superflows, which
corroborates that $^4$He superflows carry thermal
energy. This counter-intuitive heating effect bears
a phenomenological resemblance to the Peltier effect
of electric current across two distinct conductors.

\begin{figure}%
\begin{center}
\includegraphics[scale=0.2]{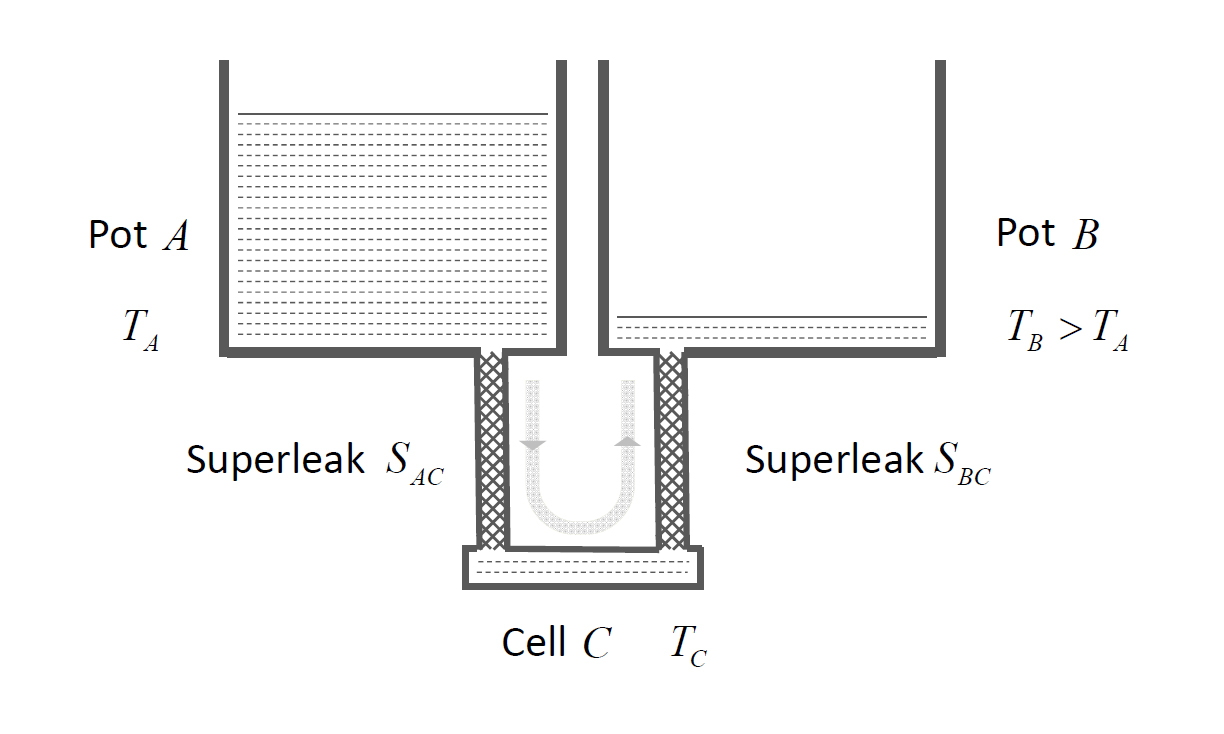}%
\caption{A schematic plot of superflow system.
} %
\label{fig:schematic}%
\end{center}
\end{figure}

The main setup of the superflow system is schematically plotted
in Fig. \ref{fig:schematic}.
The system comprises three interconnected vessels, namely, pot $A$,
pot $B$, and cell $C$, which are arranged in series using two
superleaks, labeled as $S_{AC}$ and $S_{BC}$.
Cell $C$ is thermally isolated
from its surroundings except for its thermal links to the pots via the
superleaks.

The experimental procedure involves initially filling pot $A$
with superfluid $^4$He and subsequently establishing superflows
through $S_{AC}$, cell $C$, and $S_{BC}$ by setting a
positive temperature difference between pot $B$ and $A$
({\it{i.e.}}, the fountain effect), resulting in the transportation
of superfluid from pot $A$ to $B$.
One would expect that
the temperature of cell $C$ ($T_C$) would stabilize between the temperatures
of pot $B$ ($T_B$) and pot $A$ ($T_A$). However, observations
indicate that superflows can significantly heat cell $C$, resulting
in a steady-state value of $T_C$ that exceeds $T_B$ by over
one hundred millikelvins.

\begin{figure}%
\begin{center}
\includegraphics[scale=0.15]{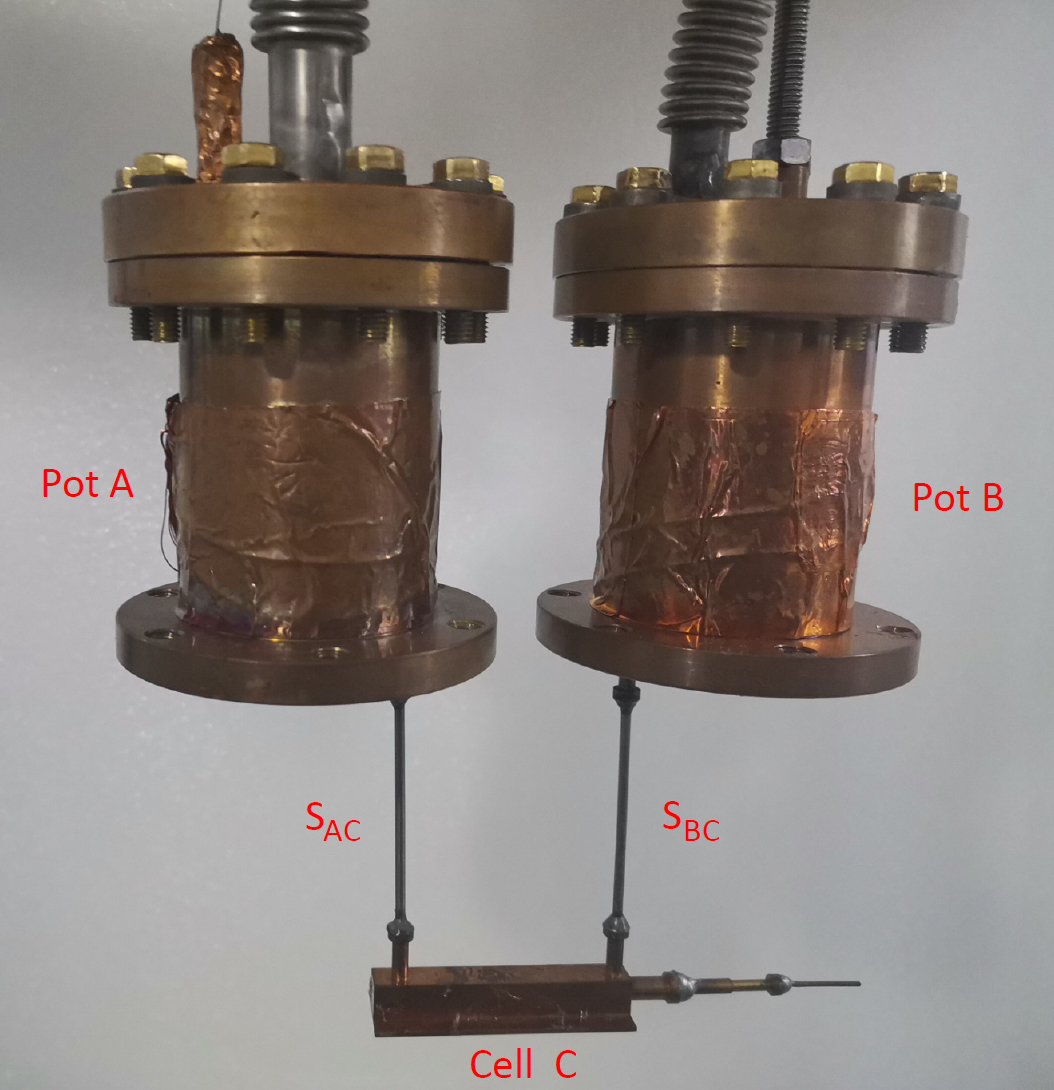}
\caption{A picture of superflow system. $S_{AC}$ and $S_{BC}$ refer to the superleaks.
} %
\label{fig:photos}%
\end{center}
\end{figure}

The experiment is carried out on a  two-stage Gifford-McMahon 
refrigerator with a cooling power of 1 $W$ at 4.2 $K$, and a 
base temperature of 2.4 $K$. To reach the superfluid temperature regime, 
a liquid $^4$He cryostat is constructed based on the scheme provided in
  Ref. \cite{field}.  The cryostat involves a stainless 
steel capillary, which has 
an inner diameter (i.d.) of 0.18 $mm$, an outer diameter (o.d.) of 0.4 $mm$,
 and a length
 of 1 $m$, which acts as the Joule-Thomson 
impedance. A copper pot with an i.d. 
of 4.0 $cm$ and a volume of 78 $cm^{3}$ is used to collect
 liquid $^4$He and serve as pot $A$ of the superflow
system.
Another identical copper pot is employed as pot $B$. 
The inner cavity of cell $C$,  made of a small 
copper block, is predominantly cylindrical with 
a diameter of 3 $mm$ and a length of 40 $mm$. Two superleaks are
 made of stainless 
steel tubes packed with jeweler's 
rouge powder (with an average particle size of 70 $nm$ determined by TEM).
The tube for   $S_{AC}$ has an i.d. of 0.8 $mm$, an o.d. of 2.0 $mm$,
and a length of 65 $mm$, while the tube for $S_{BC}$ has an i.d. of 1.0 $mm$, 
an o.d. of 2.0 $mm$
and a length of 65 $mm$.
The two superleaks are soft soldered to cell $C$, and they are positioned such
 that the lower end of each superleak  is situated near one 
end of cell $C$'s cylindrical cavity. The upper end of
  $S_{AC}$ is connected to pot $A$, and the upper end of $S_{BC}$ 
is connected to pot $B$ (see Fig. \ref{fig:photos}).

A combination of copper braids and brass strips is used 
as a thermal link to connect pot $A$ with a cooling plate 
mounted directly on the second stage of the refrigerator. 
This thermal link has a thermal conductance of 
approximately 2 $mW/K$ at 2 $K$. On the other hand, the 
main thermal link between pot $B$ and its surroundings 
is a copper braid connecting the two pots. Resistance wires 
are wrapped around both pots to serve as heaters. Pot $A$ and 
pot $B$ are fitted with pumping lines and valves to 
regulate the pumping rate, providing a further method 
for the temperature control of the pots.

Calibrated carbon ceramic resistors \cite{ccs} are used as 
temperature sensors to measure $T_A$, $T_B$, and $T_C$, 
with an accuracy of 5 $mK$ below 2.5 $K$. The 
dissipation power of the temperature sensor on cell $C$ 
is kept significantly below $10^{-7}$ $W$, ensuring 
that its heating effect is limited.

The refrigerator is equipped with two radiation 
shields, which are composed of copper and coated 
with a thin layer of nickel. These shields are 
installed in each of the two stages of the 
refrigerator, respectively. One shield operates at 
a temperature of approximately 45 $K$, while the 
other shield maintains a temperature below 2.8 $K$. 
Great care is taken to shield cell $C$ from 
exposure to thermal radiation that originates 
from sources with a temperature exceeding 3 $K$.

To initiate the accumulation of liquid $^4$He 
(with a purity of 99.999\%) in pot $A$, the 
temperature $T_A$ is deliberately raised above 
the $\lambda$ point, which prevents superflow 
through the superleak $S_{AC}$, while 
simultaneously allowing cell $C$ and pot $B$ to 
remain unoccupied. After a substantial amount 
of liquid $^4$He has accumulated in pot $A$, 
the pumping rate is adjusted to lower $T_A$. As 
$T_A$ drops below the $\lambda$ point, superflow 
is initiated through the superleak $S_{AC}$, 
filling cell $C$ in the process. Establishing 
the superfluid transport from pot $A$ to pot $B$ 
via cell $C$ can be accomplished by setting 
$T_B$ higher than $T_A$. This transport can be 
sustained for several hours. During this transport 
process, it is observed that $T_C$ remains 
stable once it reaches a steady value. Tab. \ref{tab:Ts} 
lists the various stable values of $T_C$ for 
different predetermined combinations of $T_A$ and $T_B$.

\begin{table}
\centering
\begin{tabular}[c]{|l|c|r|}
\hline 
$T_A$ ($K$) & $T_B$ ($K$) & $T_C$ ($K$) \\ \hline 
1.500(4) &  1.700(4)  & 1.847 (1) \\ \hline
1.600(4) &  1.800(4)  & 1.927 (1)\\ \hline
1.600(4) & 1.900(4)   & 2.014 (1)\\ \hline
\end{tabular}
\caption{Steady values of $T_C$ at various combinations of $T_A$ and $T_B$. The
 number in the parenthesis denotes the fluctuation amplitude of 
the measured temperature 
over several hours. $T_C$ is observed to be highly 
stable, with an deviation 
no larger than 1 $mK$. }
\label{tab:Ts}
\end{table}

\begin{figure}[htbp]
\begin{center}
\includegraphics[scale=1.0]{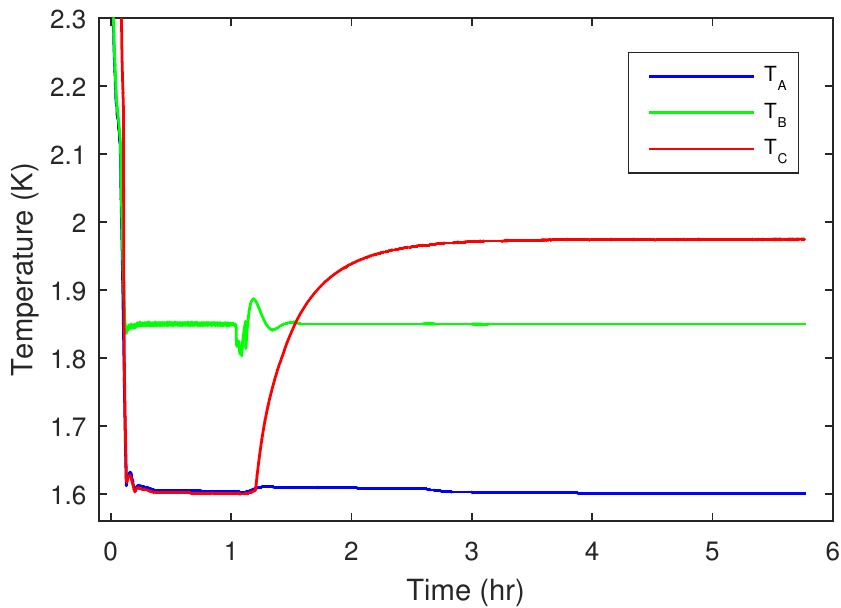}
\caption{A heating process of cell $C$. Note that after reaching a steady state, 
the temperature of cell $C$ exhibits a high degree of stability.} 
\label{fig:heating}%
\end{center}
\end{figure}

A heating process of cell $C$ is plotted in Fig.~\ref{fig:heating}.
At the initial moment presented in the figure, pot $A$ 
contains liquid $^4$He with
a temperature above the $\lambda$ point, while both pot $B$ 
and cell $C$ are empty. Thereafter, pot $A$ is pumped and 
its temperature decreases, which causes an accompanying drop 
in $T_B$ due to the relatively large thermal link between 
the two pots. At a certain point, the resistance wire 
around pot $B$ is activated, which stabilizes $T_B$ at a set value of 1.85 $K$.

As $T_A$ decreases, $T_C$ also decreases. Once both 
temperatures dip below the $\lambda$ point, superfluid 
transport between pot $A$ and cell $C$ is initiated, which 
plays a crucial role in determining the value of $T_C$ 
relative to $T_A$. $T_C$ remains a few millikelvins 
below $T_A$ during the period of superfluid filling 
of cell $C$. The difference between $T_C$ and $T_A$ 
remains relatively locked owing to a negative 
feedback mechanism. Specifically, if $T_C$ falls further 
below $T_A$, the fountain pressure of the superfluid that 
stems from the temperature difference between $T_A$ and $T_C$ 
overcomes the gravitational pull and directs the superfluid 
back from cell $C$ to pot $A$, thereby elevating $T_C$. Conversely, 
if $T_C$ approaches $T_A$, the overall force, which 
comprises both fountain pressure and gravitational pull, 
conducts the superflow from pot $A$ to cell $C$, leading 
to a reduction in $T_C$.


When the filling of cell $C$ nears completion, the 
initiation of superfluid transport between cell $C$ 
and pot $B$ causes significant fluctuations of $T_B$, 
which oscillates several times within a 
period of around thirty minutes. The underlying 
cause of these fluctuations is the relatively 
small heat capacitance of the initially empty 
copper pot $B$. The introduction of even a small 
amount of cold superfluid $^4$He into pot $B$ 
can provoke a considerable variation of $T_B$, 
and the temperature stabilization system 
cannot respond in a timely manner. As a 
certain amount of superfluid $^4$He accumulates 
in pot $B$, subsequent injections of cold 
superfluid no longer produce a dramatic change 
in $T_B$, thus enabling the temperature 
stabilization of $T_B$ to be restored.

Once the transport of superfluid $^4$He through 
these two superleaks is fully established, $T_C$ 
increases steadily and reaches a value 120 $mK$ 
above $T_B$, which demonstrates a counter-intuitive 
heating phenomenon. The data presented in 
Fig. \ref{fig:heating} allows for useful 
estimations of the heating powers received by 
cell $C$ at two distinct stages. During the 
first stage, which takes approximately an 
hour and corresponds to the filling of 
superfluid into cell $C$, the cell receives 
heat transfer from pot $B$ via thermal 
conduction while being cooled by the influx 
of superfluid (note that this heat transfer 
drives superfluid to flow into cell $C$). The 
heat received by cell $C$ during this stage 
corresponds to the thermal energy of 
liquid $^4$He with a volume of 0.1 $cm^3$ 
at a temperature of 1.6 $K$, resulting 
in a heating power denoted by $P_{stage1}$ 
of approximately 1.5 $\mu W$. Right after 
the first stage, the temperature of cell $C$ 
(superfluid filled) rises from 1.6 $K$ to 
1.85 $K$ over a duration of 20 minutes, which 
is considered as the second stage. The heat 
received during this stage corresponds to roughly 
the change of thermal energy of liquid $^4$He 
in the cell, resulting from the temperature change 
from 1.6 $K$ to 1.85 $K$. The heating power 
denoted by $P_{stage2}$ is 7.3 $\mu W$.
 


It is worth noting that even if cell $C$ could 
experience some abnormal background heating during 
the experiment, this heating should also be 
present in the first stage as well and thus 
has a smaller power than $P_{stage1}$. Based on 
the fact that $P_{stage2} > 4 P_{stage1}$, it 
is concluded that the major portion of $P_{stage2}$ 
is sourced from the superflows rather than 
some abnormal heating background. The superflows 
remain responsible for the heating of cell $C$ for 
the rest of the time presented in Fig. \ref{fig:heating}. 
Since the superflows' kinetic energies are 
negligible, it can be inferred that the heating 
resulting from the superflows must originate 
from their thermal energies.

The observed heating phenomenon of superflows is 
someway analogous to the Peltier effect, which 
occurs when an electric current flows across 
two distinct conductors. Specifically, in this 
phenomenon, the thermal energy density of 
the inlet superflow entering cell $C$ is larger 
than that of the outlet superflow exiting    
the cell, resulting in the heating of the 
cell. Fundamentally, the difference between 
the inlet and outlet superflows' thermal energy 
densities is attributed to the velocity dependence 
of a superflow's thermal energy.

The velocity behavior of the superflows in the two 
superleaks is complex, as they are frictionless and 
lack a means of stabilizing their velocities. The superflow 
in superleak $S_{AC}$ undergoes continuous acceleration 
or deceleration due to the pressure difference between 
superfluid in pot $A$ and that in cell $C$, wherein the 
fountain pressure, related to $T_C - T_A$, comprises a 
significant part of the overall pressure difference. The 
pressure in cell $C$ can increase abruptly as it 
transitions from a nearly full state to a fully filled state, 
which regulates the velocity of inlet superflow and 
prevents it from attaining high velocity. On the 
other hand, the pressure increase in cell $C$ can 
accelerate the outlet superflow, causing it to reach 
a large velocity regime. Furthermore, if $T_C > T_B$, 
the superflow in $S_{BC}$ can reverse its flow 
direction when cell $C$ deviates from a fully 
filled state, leading to a shorter actual flow time 
of the outlet superflow than that of the 
inlet superflow. These analyses indicate an 
asymmetry between the velocity distributions of the 
inlet and outlet superflows, resulting in a disparity 
in their thermal energies.

To corroborate our findings, we conducted a 
comparative experiment in which we replaced the 
superleak $S_{BC}$ by a solid rod of stainless 
steel of similar dimensions (radius of 2.0 $mm$ and 
a length of 65 $mm$). This experiment only allows 
superfluid transport between pot $A$ and cell $C$ while 
it is obstructed between cell $C$ and pot $B$. As 
anticipated, and confirmed, the steady value of $T_C$ 
is well between $T_A$ and $T_B$. The comparison provides 
additional robust evidence that the unusual heating 
phenomenon is not due to some abnormal background heating.


In earlier experiments, it appears that $^4$He superflows 
carry negligible thermal energies. This observation is 
commonly attributed to the absence of controlled 
superflow velocities, which often approach the critical 
value during experimentation. However, in contrast 
to those experiments, the present investigation 
involves partial control of the superflow velocities, 
leading to the observation of an intriguing heating phenomenon.
 
\section*{VI. Conclusions and perspectives}
Superfluid $^4$He is one of the most fascinating systems
in condensed matter physics, and it continues to unveil
surprising quantum behaviors more than eight decades
after the discovery of superfluidity. Despite substantial
efforts in the past, however, the exact
microscopic understanding of this system had remained elusive
owing to the complexity of the many-body quantum problem.

This paper presents a fundamental property of the quantum
system of superfluid $^4$He, specifically, the grouping behavior
of its low-lying levels. Based on this observation, we
offer a natural explanation of superfluidity. Furthermore, we
demonstrate that the group-specific nature of the thermal
equilibrium state of a superflow leads to another
intriguing feature of this system: the thermal energy density
of a superflow at a given temperature depends on its velocity,
with higher velocities leading to lower thermal energy density. This
fundamental connection between the thermal and hydrodynamic
properties of the system plays an essential role in
various phenomena, such as the mechano-caloric effect.
We also provide a natural explanation
for the presence of dissipation in a superflow undergoing
time-varying motion, in contrast to steady motion. In addition,
we suggest that several condensed matter systems, which exhibit
interesting low-temperature phases described by an order
parameter, share the same common microscopic basis: their low-lying
quantum states are grouped. Each group of quantum levels can
produce a distinct macroscopic thermal state of the system with a
particular value of the order parameter.

Drawing inspiration from the theoretical
findings, we conducted an experiment to
investigate $^4$He
superflows, which led to the observation
of a counter-intuitive heating phenomenon. Our
results confirm that a superflow can carry
considerable thermal energy, related
to its flow velocity.

The theoretical picture of superfluid $^4$He
presented in this paper has the potential to shed light on
some other intriguing phenomena that have
not yet been fully elucidated in the system.
Additionally, it can lead to predictions of
a couple of novel behaviors which
can be investigated empirically. A thorough integration
of theoretical and experimental aspects in this field
can be pursued based on this picture, which will
reinforce the conviction
that quantum mechanics offers a definitive
account of laboratory systems.


\begin{thebibliography}{10}
\bibitem{finitesize} Practically, any sample of superfluid $^4$He in a laboratory 
has a finite size. However, the system with $N \to \infty$ in the 
thermodynamic limit can be obtained from a finite-sized system 
through appropriate scaling, which preserves its inherent physical properties.
\bibitem{kinetic} The average rate of transfer of superfluid $^4$He 
from pot $A$ to pot $B$ ranges between 3 to 4 $cm^3$ per hour. 
The flow velocities through superleaks are typically a few millimeters 
per second. The associated kinetic energy per $^4$He atom is
 approximately $10^{-29}$ $J$, corresponding to a temperature scale of 1 $\mu K$, 
which is negligible for a $^4$He system operating above 1.6 $K$. Furthermore,
 as discussed later in the manuscript, the inlet superflow exhibits a 
smaller velocity than the outlet superflow. Consequently, any change 
in the kinetic energy of the superflows will result in a minor 
cooling effect on cell $C$ rather than a heating effect.
\end{thebibliography}
\end{document}